\documentclass{article}

\usepackage{arxiv}
\usepackage{graphicx}
\usepackage{authblk}

\usepackage[utf8]{inputenc} 
\usepackage[T1]{fontenc}    
\usepackage{hyperref}       
\usepackage{url}            
\usepackage{booktabs}       
\usepackage{amsfonts}       
\usepackage{nicefrac}       
\usepackage{microtype}      
\usepackage{lipsum}
\usepackage{color}

\title{Quantifying the performance of machine learning models in materials discovery}

\author[1]{Christopher K. H. Borg}
\author[1]{Eric S. Muckley}
\author[1]{Clara Nyby}
\author[1]{James E. Saal}
\author[2]{Logan Ward}
\author[3]{Apurva Mehta}
\author[1]{Bryce Meredig}

\affil[1]{Citrine Informatics, Redwood City, CA, United States}
\affil[2]{Argonne National Laboratory, Lemont, IL, USA}
\affil[3]{Stanford Synchrotron Radiation Lightsource, SLAC National Accelerator Laboratory, Menlo Park, California 94025, USA}

\affil[*]{corresponding author(s): James E. Saal (jsaal@citrine.io)}

\begin{document}
\maketitle

\begin{abstract}
The predictive capabilities of machine learning (ML) models used in materials discovery are typically measured using simple statistics such as the root-mean-square error (RMSE) or the coefficient of determination ($r^2$) between ML-predicted materials property values and their known values. A tempting assumption is that models with low error should be effective at guiding materials discovery, and conversely, models with high error should give poor discovery performance. However, we observe that no clear connection exists between a “static” quantity averaged across an entire training set, such as RMSE, and an ML property model’s ability to dynamically guide the iterative (and often extrapolative) discovery of novel materials with targeted properties. In this work, we simulate a sequential learning (SL)-guided materials discovery process and demonstrate a decoupling between traditional model error metrics and model performance in guiding materials discoveries. We show that model performance in materials discovery depends strongly on (1) the target range within the property distribution (e.g., whether a 1st or 10th decile material is desired); (2) the incorporation of uncertainty estimates in the SL acquisition function; (3) whether the scientist is interested in one discovery or many targets; and (4) how many SL iterations are allowed. To overcome the limitations of static metrics and robustly capture SL performance, we recommend metrics such as Discovery Yield ($DY$), a measure of how many high-performing materials were discovered during SL, and Discovery Probability ($DP$), a measure of likelihood of discovering high-performing materials at any point in the SL process.
\end{abstract}

\section{Introduction}
As machine learning (ML) tools become more accessible to materials researchers, utilizing ML models to design experiments is becoming commonplace.
Many of the recent successes in applying ML for materials discovery have been captured in a review by Saal et al.~\cite{saal2020machine}.
One approach to materials discovery via ML is to train an ML model (or an ensemble of models) on a property of interest, make predictions on unknown materials, and conduct a validation test, usually by experiment~\cite{meredig2014combinatorial, oliynyk2016high, zhuo2018identifying, rickman2019materials, wu2019machine, gomez2016design}. 
These methods rely on having large enough training sets that new predictions represent interpolations within explored regions.
In contrast, when training data is scarce or extrapolation is necessary, a sequential learning (SL), sometimes referred to as active learning, approach can be employed~\cite{ling2017high, bassman2018active, lookman2019active, del2020assessing, coley2020autonomous, montoya2020autonomous}.
Sequential learning involves training an initial ML model, selecting optimum candidates based on an acquisition function, verifying those predictions with simulations or experiments, and then updating the model with new data.
This iterative sequential learning loop offers an efficient means of exploring large design spaces, reducing the number of experiments necessary to realize a performance goal~\cite{nikolaev2016autonomy, lookman2016perspective, kim2019active, antono2020machine}.


Materials discovery offers unique challenges as an ML application area.
For example, materials of interest often exhibit properties with extreme values, requiring ML models to extrapolate to new regions of property space.
This challenge, and methods to address it, have been discussed previously~\cite{meredig2018can, kauwe2020can}.
Another challenge is representing a material suitably for input to an ML algorithm, either by incorporating domain knowledge, or learning representations from data.
Chemical composition-based features~\cite{ward2016general, jha2018elemnet}, have become widely used in materials discovery, but it is likely that further headroom exists for optimization of materials representations.
Finally, many materials informatics applications suffer from lack of data.
While there have been many large scale data collection efforts~\cite{blaiszik2016materials, draxl2018nomad, o2016materials}, researchers often extract data by hand to build informative training sets~\cite{nyby2021electrochemical, borg2020expanded, iwasaki2019machine, balachandran2018experimental, min2018machine, hatakeyama2019synthesis, wen2019machine}, which is a highly time-consuming process.
These unique challenges motivate the need for greater insight into the potential for success in a given SL-driven materials discovery effort.

Typically, the performance of ML is measured by the improvement in predictive capability of a model, using accuracy metrics such as root-mean-square error (RMSE) and the coefficient of determination $r^2$.
While these metrics provide robust estimates for predictive capabilities against a defined test set, their connection to the ability of SL to identify new, high-performing materials is unclear.
In recent studies, the number of experiments necessary to identify a high-performing material has been used as a metric for monitoring SL performance~\cite{ling2017high,kim2019active,xue2016accelerated}. 
Modeling benchmark datasets and tools, such as Olympus~\cite{hase2021olympus} and MatBench~\cite{dunn2020benchmarking}, have started to standardize assessment of model and dataset performance.
Notably, a recent study by Rohr et al.~\cite{rohr2020benchmarking} considers additional metrics that quantify SL performance relative to a benchmark case (typically random search). Rohr et al. focuses their study on identifying top 1\% materials from high-throughput electrochemical data and subsequent research expands on this work to compare performance across a variety of SL frameworks and datasets~\cite{palizhati2021multi, liang2021benchmarking}.
Here, we build upon these works by investigating SL performance metrics for different design problems and specific targets within those design problems. We compare our approach to Rohr et al. in more detail in Section \ref{subsection:sl_metrics}.

In this work, we explore in more detail the topic of SL performance metrics, generalizing conclusions across multiple datasets and design target ranges.
In particular, we identify a decoupling between traditional model error metrics and a model's ability select high-performance materials. 
We look at three SL performance metrics: Discovery Acceleration Factor ($DAF_n$), the average number of SL iterations required to identify $n$ materials in a target range, Discovery Yield ($DY(i)$), the number of materials in a target range identified after $i$ SL iterations (normalized by the total number of of materials in the target range), and Discovery Probability ($DP(i)$), the average number of targets found at a given SL iteration, $i$. 
Each metric is focused on the ability of an SL strategy to identify high-performance materials, rather than the error associated with model predictions. 
We then demonstrate use of these metrics with a simulated SL pipeline using a commonly available band gap database~\cite{dunn2020benchmarking}. 
Next, we focus on the challenge of ML-driven design of thermoelectric (TE) materials. 
Fundamental TE properties (Seebeck coefficient, electrical conductivity, and thermal conductivity) were extracted from the Starrydata platform~\cite{katsura2019data} and used to compute TE figures of merit (ZT and $\sigma_{E0}$)~\cite{katsura2019data}.
The same simulated SL pipeline was used for these new datasets and performance metrics were compared to identify the optimal design strategies for TE materials discovery from existing data.
We then compare the SL performance metrics and traditional model error metrics to identify general trends across multiple materials domains and compare these results to prior work.

\section{Methods}
\label{sec:methods}
A typical SL process for materials discovery begins with an initial training set of material compositions and their properties.
A ML model is trained on the initial training set and used to make predictions on a set of compositions not in the training set (known as the candidate pool or design space).
An acquisition function (detailed in Section~\ref{subsection:af}) is used to select the optimum next experiment to perform from materials in the design space.
Those experiments are then performed (either physically or by some physics-based simulation) and the results are added to the training dataset.
The improved training set is then used in place of the initial training set and the process is repeated until the design goals have been realized.

To initialize simulated sequential learning, the SL pipeline must be supplied with a set of user-defined SL configuration parameters, a dataset that contains a set of inputs (typically chemical composition) and a property to be optimized.
For a given dataset, chemical compositions were featurized with the Magpie elemental feature set~\cite{ward2016general} implemented via the element-property featurizer in Matminer~\cite{ward2018matminer}.
For all discovery tasks, random forests were employed with uncertainty estimates (estimated by calculating the standard deviation in predicted values across estimators) and paired with an acquisition function to enable the selection of a candidate from the design space, as detailed below.


\subsection{Simulated sequential learning pipeline}
\begin{figure}
    \centering
    \includegraphics[width=\textwidth]{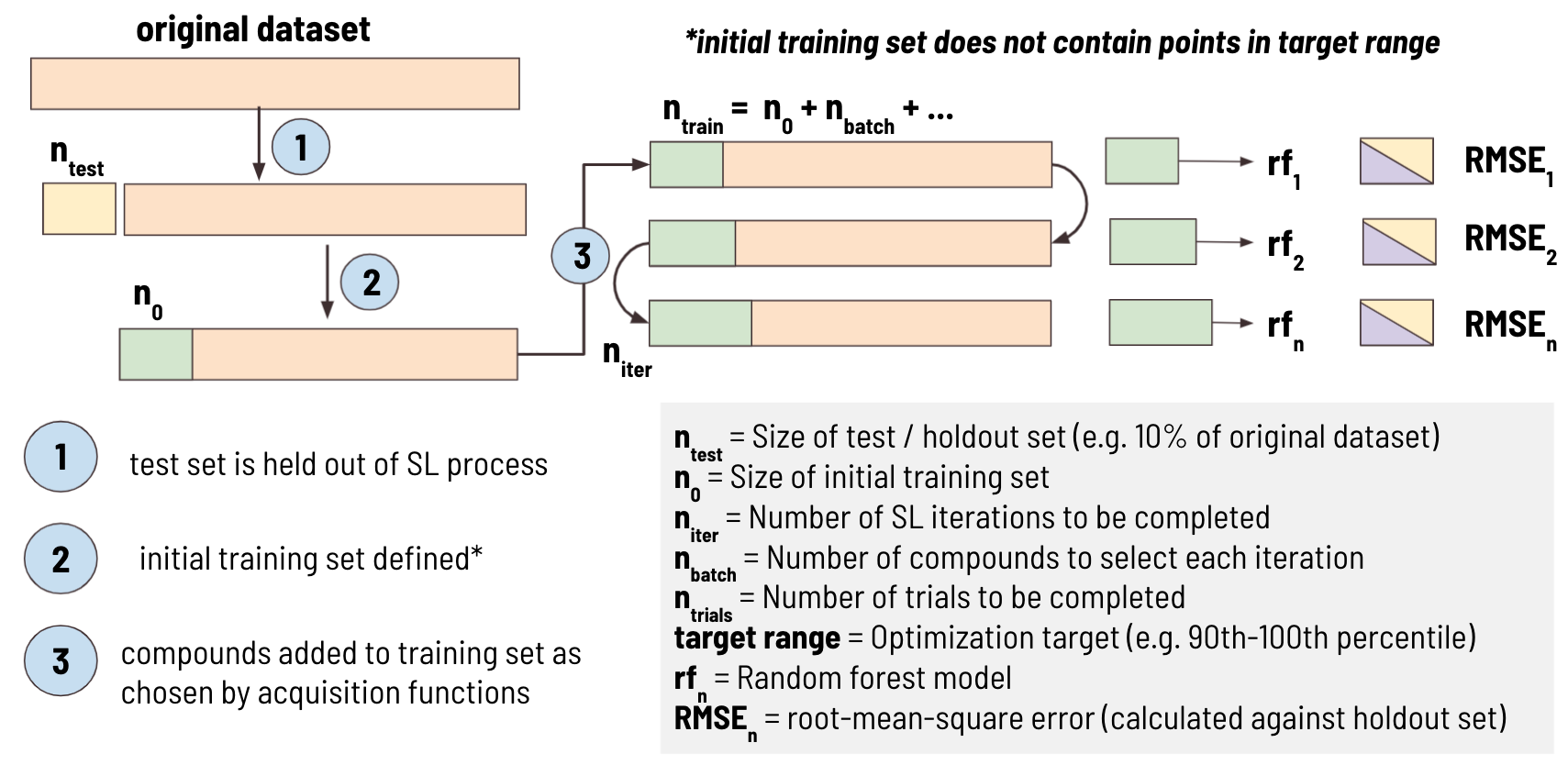}
    \caption{The sequential learning workflow used to calculate parameters of interest. A holdout dataset (n$_{test}$) is defined prior to initializing the SL process. This holdout set (denoted in yellow) is used to calculate RMSE against predicted values calculated using a model trained on the updated training set (denoted in purple) at each SL iteration. The actual training set is denoted in green and untested candidates (i.e., the candidate pool) are denoted in orange.} 
    \label{fig:sl-flow}
\end{figure}

In this work, we developed a standard processing pipeline to simulate sequential learning, a workflow summarized in Figure \ref{fig:sl-flow}.
First, 10\% of the dataset ($n_{test}$) is randomly sampled and held out of the SL process entirely.
This "holdout set" is used to calculate RMSE against the training set at each iteration in the SL process. 
This is done to ensure that the model is tested on the same test set at each iteration.
A target range is then selected comprising one of the 10 deciles of the remaining dataset (e.g., "1st decile" indicates that the SL design goal is to find materials between 0th and 10th percentile of the entire dataset).
Then, an initial training set, with size $n_0$, is sampled such that it does not contain any material in the target range.
For example, when the target range was defined as 10th decile, compounds with the highest 10\% of values would be excluded from being in the initial training set, ensuring that they are left in the candidate pool.
This was done to simulate a real-world materials discovery problem, as a typical goal is to find materials with performance superior to what is currently known. 
Finally, sequential learning is performed for $n_{iter}$ iterations using the acquisition functions (detailed in Section~\ref{subsection:af}) to find materials as close as possible to the mean of the target range.
At each iteration, $n_{batch}$ compound(s) are added to the training set, SL metrics are calculated (defined in Section~\ref{subsection:sl_metrics}), and
the entire process is repeated $n_{trials}$ times to determine the trial-to-trial stochastic uncertainty of the SL pipeline.

In this work, the following sequential learning configuration was used: $n_{test}$ size = 10\% of dataset, $n_0$ = 50, $n_{iter}$=100, $n_{batch}$ = 1, $n_{trials}$=100.
This configuration was thought to be well-aligned with traditional materials discovery problems.
For example, initial training sets were limited to 50 points, as it is expected many experimental studies wishing to employ ML would have at least this number of datapoints.
In our SL workflow we set $n_{batch}$ = 1 as experiments are often performed one at a time, however, this parameter is adjustable to values greater than 1.
By design, the first step in the SL process is the selection of the training set (as opposed to selecting training set points and then conducting a round of SL); therefore, an SL workflow where $n_{iter}$=100 has a single step to select training set points followed by 99 rounds of SL.

\subsection{Acquisition functions}
\label{subsection:af}
Four acquisition functions are considered in this work: expected value (EV),  which selects candidates whose predicted property value lies closest to the mean of the target window; expected improvement (EI), which selects candidates that are likeliest to fall within the target window based on predicted property values and the uncertainties in those predictions; maximum uncertainty (MU), which selects the candidate with largest prediction uncertainty; and random search (RS)~\cite{Garnett2015}, where a candidate is selected at random.
EV is an exploitative acquisition function that tends to locally improve upon known good materials. MU, in contrast, is meant to be more exploratory, focusing on model improvement by selecting candidates with the most uncertain prediction.
EI is a hybrid approach that attempts to improve materials performance while also taking uncertainty into account.
RS is included here for comparison, with the intuition that a directed acquisition function should outperform random selection in SL.
The comparison of these functions seeks to demonstrate tradeoffs made when considering the uncertainty associated with a ML prediction and balancing exploration and exploitation.

\subsection{SL figures of merit} 
\label{subsection:sl_metrics}
To measure the performance of the simulated SL efforts, we used the following metrics: 
\begin{enumerate}
    \item \textbf{Discovery acceleration factor ($DAF_n$)}: The average number of SL iterations required to identify $n$ compounds in the target range, relative to random search. For example, if RS takes 10 iterations to identify a target compound and EV takes 5 iterations, then $DAF_1$(EV) = 2, indicating that EV identifies a target compound twice as fast as RS. The number of iterations for RS for $DAF_1$, $DAF_3$, and $DAF_5$ were estimated to be 10, 30, and 50 respectively as each target range is comprised of a decile of the dataset. Rohr et al.~\cite{rohr2020benchmarking} proposed "acceleration factors" to refer to the number of iterations needed to achieve a particular SL performance goal (e.g., a desired DY or DP value), normalized by the number of iterations needed for RS to accomplish the same goal. Acceleration Factor (AF) is also more broadly defined as the "reduction of required budget (e.g. in time, iterations, or some other consumed resource) between an agent (model + acquisition function) and a benchmark case (e.g. random selection) to reach a particular fraction of ideal candidates"~\cite{palizhati2021multi}. In our case, $DAF_n$ is specific to the number of iterations to find $n$ target materials relative to random search.
    
    The equation for $DAF_n$ is given by:
    \begin{equation}\label{eqn:daf}
    \mathrm{DAF_n} = \frac{i_n(EV, EI, MU)}{i_n(random)}
    \end{equation}
    where $i_n(EV, EI, MU)$ = the number of SL iterations it takes to find n target compounds via EV, EI, or MU and $i_n(random)$ = the number of SL iterations it takes to find n target compounds via random search. 
    
    \item \textbf{Discovery yield ($DY(i)$)}: The number of compounds in the target range identified after a set number of iterations divided by the total number of compounds in the target range. For example, the band gap 10th decile range is comprised of 193 compounds (after removing the holdout set). On average, after 20 iterations, $DY_{i=20}$(EI) =  0.07$\pm$0.02; indicating $\approx$7\% of targets were discovered. This is meant to represent, for a given number of experiments, how many high-performing compounds a researcher could expect to find. Rohr et al.~\cite{rohr2020benchmarking} proposed $^{all}$ALM$_i$ as the "fraction of the top
percentile catalysts that have been measured by cycle $i$". We interpret $^{all}$ALM$_i$ to be an equivalent figure of merit to $DY(i)$ but applied to identifying top 1\% materials (rather than a decile window).
    
    The equation for $DY(i)$ is given by:
    \begin{equation}\label{eqn:dy}
    \mathrm{DY(i)} = \frac{1}{t_{total}}t_i
    \end{equation}
    where $t_i$ = number of targets found by iteration $i$, and $t_{total}$ = total targets in dataset.

    \item \textbf{Discovery probability ($DP(i)$)}: The likelihood of finding a target at a given SL iteration. For example, as shown in Figure~\ref{fig:bandgap_dp}, after one iteration of using EI to identify 10th decile band gap materials, $DP$(1)(EI)=0.4 (i.e., 40 out of 100 trials identified a target compound after 1 SL iteration). In contrast, after 99 iterations, $DP$(99)(EI)=0.6 (i.e., 60 out of 100 trials identified a target compound after 99 SL iterations). This is meant to estimate the likelihood of identifying a target compound at every point in the SL process, independent of previous SL cycles. In contrast to $DY(i)$, $DP(i)$ may increase or decrease with increasing SL iterations.

    The equation for $DP(i)$ is given by:
    \begin{equation}\label{eqn:dp}
    \mathrm{DP(i)} = \frac{1}{n_{trials}} \sum_{n=1}^{n_{trials}} TF_{i} \{0,1\}
    \end{equation}
    where $TF_{i}$ = target found boolean (1 if target was found, 0 if not found) at iteration $i$ and $n_{trials}$ = number of trials.
    
    Consequently, $DP(i)$ is also the derivative of $DY(i)$ with respect to iterations when correcting for the total number of targets in the dataset. The equation is given by:
    
    \begin{equation}\label{eqn:dp2}
    \mathrm{DP(i)} = \frac{1}{t_{total}} \frac{dDY(i)}{di}
    \end{equation}
    where $t_{total}$ = total targets in dataset.
    
    \item \textbf{Root-mean-square error ($RMSE$)}:  The standard deviation of the residuals calculated from the difference between predicted and actual values of compounds in the holdout set (i.e., compounds removed from the dataset before SL process is initiated). The RMSE is given by 

    \begin{equation}\label{eqn:rmse}
    \mathrm{RMSE} = \sqrt{\frac{1}{n} \sum_{i=1}^{n} \left| y_i - \hat{y}_i \right|^2}
    \end{equation}

    where $n$ is the number of samples and $y_i$ and $\hat{y}_i$ are the $i$th
    actual and predicted value of the test set, respectively.

    \item \textbf{Non-dimensional model error ($NDME$)} = RMSE normalized by the error of guessing the mean value (GTME) of the training set for each holdout test point. NDME is a means to directly compare model accuracy across different properties. This guess the mean error (GTME) is defined as
    
    \begin{equation}\label{eqn:gtme}
    \mathrm{GTME} = std(y_{holdout\_set})
    \end{equation}
    
    and the NDME is then
    \begin{equation}\label{eqn:ndme}
    \mathrm{NDME} = \frac{\mathrm{RMSE}}{std(y_{holdout\_set})}
    \end{equation}

    1-NDME$^2$ is approximately equivalent to the coefficient of determination ($r^2$). However an important difference between the two metrics is that NDME is normalized by the variance of a sample of a dataset (in this case, the holdout set), whereas $r^2$ is normalized by the variance of an entire dataset.
    \end{enumerate}

\subsection{Datasets}\label{datasets}
To calculate baseline SL metrics against known data, we utilized a dataset of measured band gaps from the MatBench repository \cite{dunn2020benchmarking}. Specifically, we attempted to predict band gap from a set of inorganic compounds in the MatBench v0.1 test dataset ("matbench\_expt\_gap"). The original report ~\cite{zhuo2018predicting} notes that duplicate records (records with the same chemical composition) were removed by assigning values based on the closest experimental value to the mean of all reports for that composition. In addition to this, we removed all zero band gap records, resulting in a dataset size of 2,154 records.

Properties for thermoelectric materials were extracted from the Starrydata2 thermoelectrics database~\cite{katsura2019data}.
New records are uploaded daily and at the time of extraction (August 9th, 2021), data from 34,738 samples were retrieved.
Due to the nature of how thermoelectric properties are reported in literature, a TE data extraction and property calculation pipeline was developed for this work.
This pipeline can be used to extract any number of TE properties available in Starrydata2 at arbitrary temperatures. For sparse records, properties were calculated when possible (e.g., if a given record was missing ZT but the Seebeck coefficient, electrical conductivity, and thermal conductivity were given, ZT was calculated from those extracted values).
For all properties, if there was more than one reported value for a particular composition, we took the mean-value to be the ground-truth.
Additionally, we focus on "111-type" compounds, defined as materials containing at least 3 metallic elements in ratios close to that of 1:1:1.
This resulted in dataset sizes of 626, 618, and 705 for ZT, $\kappa_{total}$, and $\sigma_{E0}$ respectively.
More details of the TE data extraction and property calculation pipeline are given in the supplementary information.

\section{Results}
\subsection{Band gap Benchmark}
\begin{figure}
    \centering
    \includegraphics[width=\textwidth]{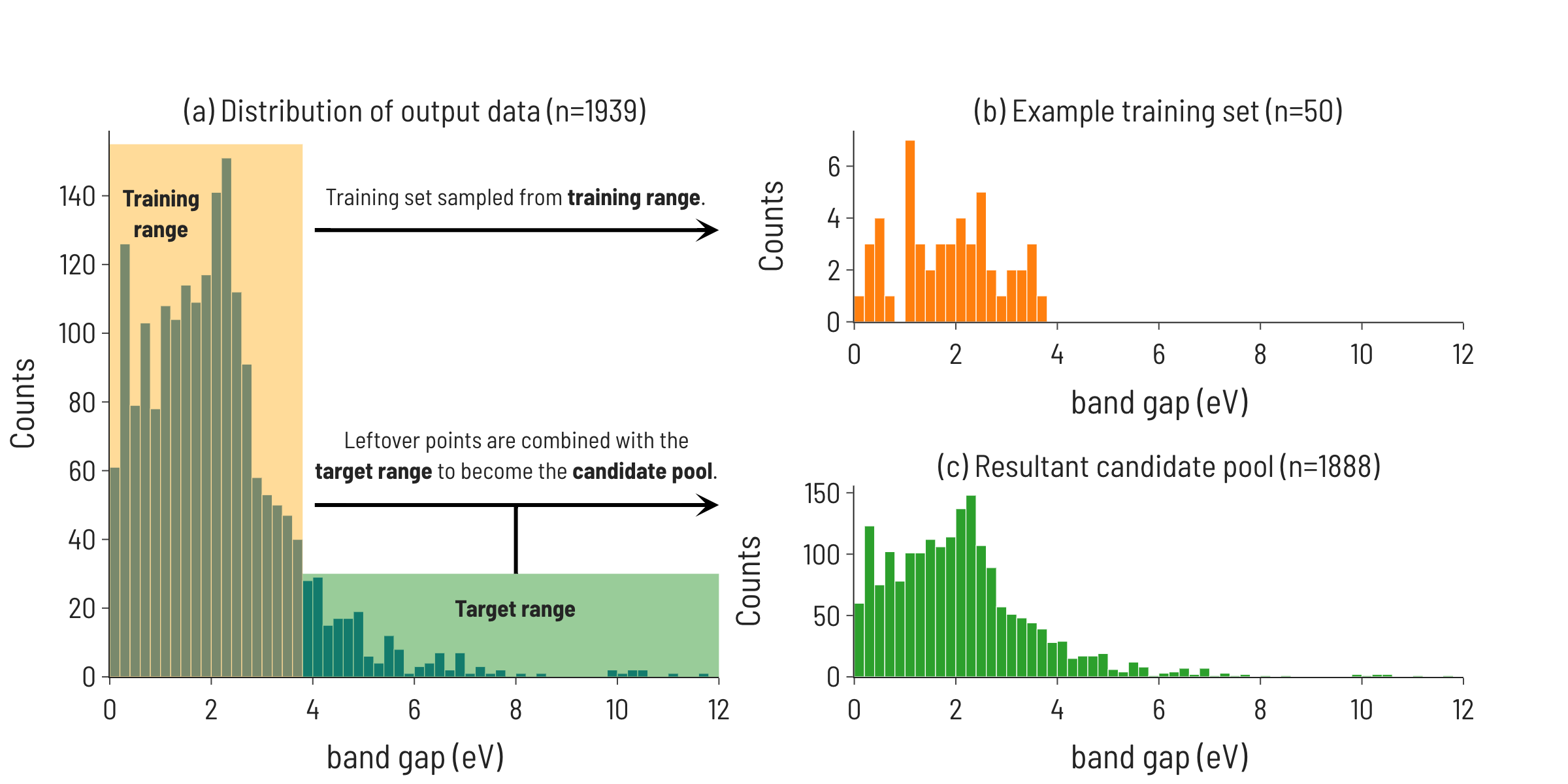}
    \caption{(a) Histogram of experimental band gaps of crystalline compounds from the MatBench benchmark dataset\cite{dunn2020benchmarking}, with the training and target ranges highlighted for the 10th decile target range. Note: this is the distribution of points after holding out a random 10\% of the dataset to be used as a test set (n = 2154 - 215 = 1939). (b) Distribution of training set sourced from points outside the target range (n = 50). (c) The resulting candidate pool of points not selected for training and points in the target range (n = 1889).
    } 
    \label{fig:bandgap_config}
\end{figure}

We begin our analysis by considering the discovery of materials with particular values of band gaps using the Matbench dataset.
The initialization of the SL workflow and configuration paramaters for the data are summarized in Figure \ref{fig:bandgap_config}. 
The distribution of band gap values is shown in Figure \ref{fig:bandgap_config}a, with the training and target regions highlighted. 
An example training set, sampled from within the training range, and the resulting candidate pool consisting of the remainder of the dataset, are shown in Figure~\ref{fig:bandgap_config}b and Figure~\ref{fig:bandgap_config}c, respectively.

\begin{figure}
    \centering
        \includegraphics[width=\textwidth]{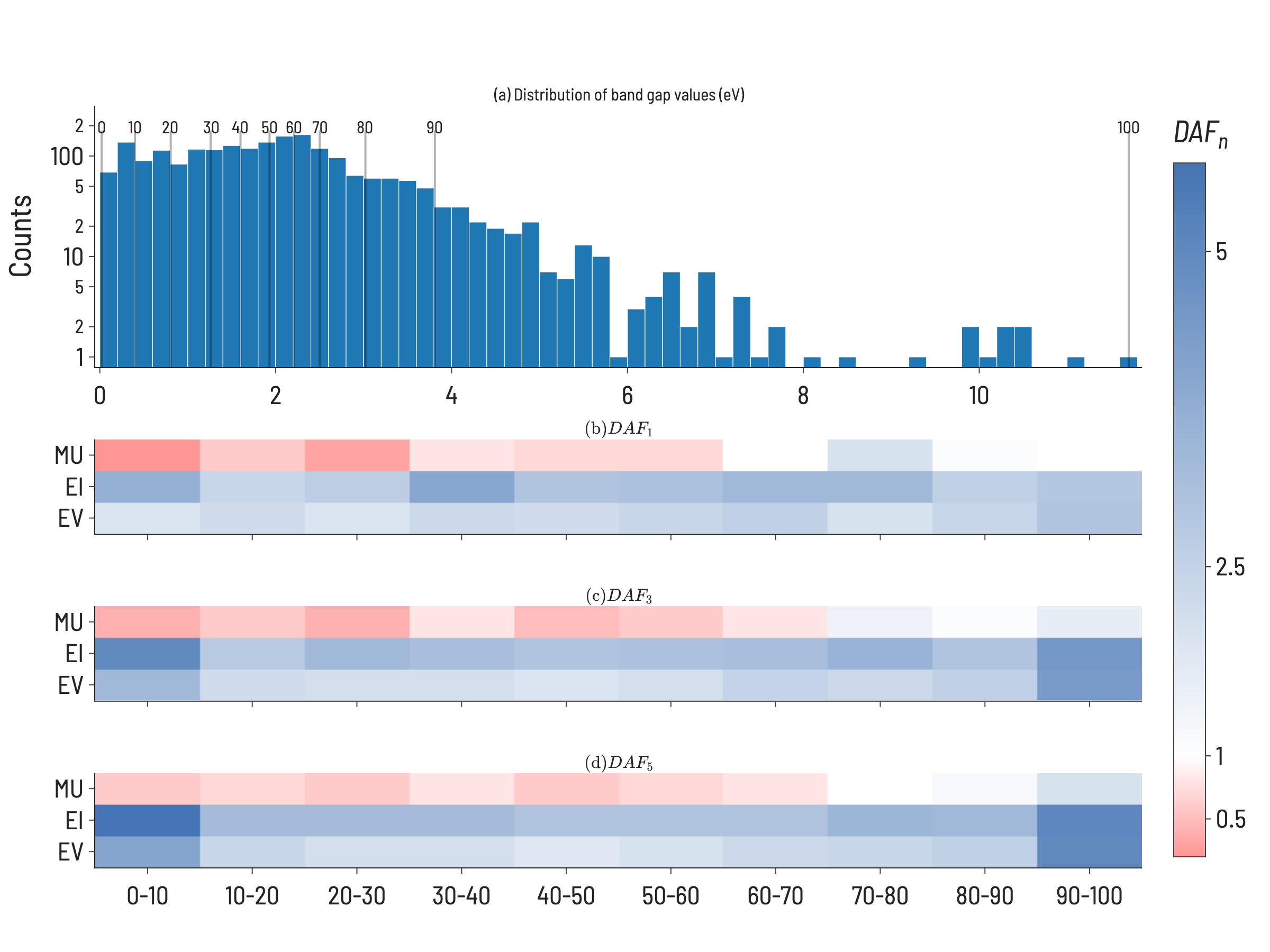}
    \caption{(a) Distribution of band gap data sourced from MatBench. Vertical lines are shown to illustrate the range of values incorporated in each decile. (b-d) Discovery acceleration factor ($DAF_n$) to the first 1, 3, or 5 discoveries: The target deciles are shown along the x-axes and the acquisition functions are shown along the y-axes. A diverging color scheme is used to indicate performance (blue = the target compound(s) is(are) identified faster than RS, red = the acquisition function is outperformed by RS for the given target). For example, $DAF_3$ shows iterations to identify 3 compounds in the target decile relative to RS (30 iterations). $DAF_n$ values were averaged over all trials ($n_{trials}$=100).} 
    \label{fig:bandgap_heatmap}
\end{figure}

The discovery acceleration factor ($DAF_n$) to find 1, 3, and 5 compounds for every combination of decile target range and acquisition function were recorded.
The results are shown in Figure~\ref{fig:bandgap_heatmap}, where $DAF_n$ is presented as a heatmap such that $DAF_n$ > 1 (faster than RS) ranges from white to blue and $DAF_n$ < 1 (slower than RS) ranges from white to red.
$DAF_n$ = 1 implies that the acquisition function identifies $n$ targets at a rate equal to that of RS.
Our results in Figure~\ref{fig:bandgap_heatmap} demonstrate that the performance of each acquisition function is strongly dependent on the target decile within the property distribution.
For example, EI performs particularly well compared to RS when targeting multiple very high and very low band gap values ($DAF_5$ = 5.6 and 5.0 for 1st and 10th deciles, respectively).
However, its advantage relative to RS is much smaller when targeting intermediate band gap values ($DAF_5$= 2.9 and 3.0 for 5th and 6th deciles, respectively).
Figure~\ref{fig:bandgap_heatmap} also shows that, as one might anticipate, the performance advantage of ML-guided discovery increases for extreme values in the property distribution after finding the first such extreme value.
This increase is clearest at the high and low extreme value regions, 1st and 10th decile materials, where EI and EV see a 2-fold increase in performance from identifying a single target material ($\approx$2x faster than RS) to identifying 5 target materials ($\approx$4x faster than RS).  
Interestingly, despite these two target ranges having different distributions in the dataset (as shown in Figure~\ref{fig:bandgap_heatmap}a), there is a similar acceleration to reach 3 or 5 target compounds, implying that the property value width of a target decile range is not correlated to difficulty in identifying materials in that range.
MU, regardless of the target decile, is the slowest to finding target compounds, and in deciles below the 8th decile, is slower than RS, implying a design strategy based only on uncertainty estimates would be unsuccessful. 
However, EI appears to be quicker than EV across all target ranges, indicating that the uncertainty-aided acquisition function can generally hasten discovery activities, at least for this dataset.

\begin{figure}
    \centering
    \includegraphics[width=\textwidth]{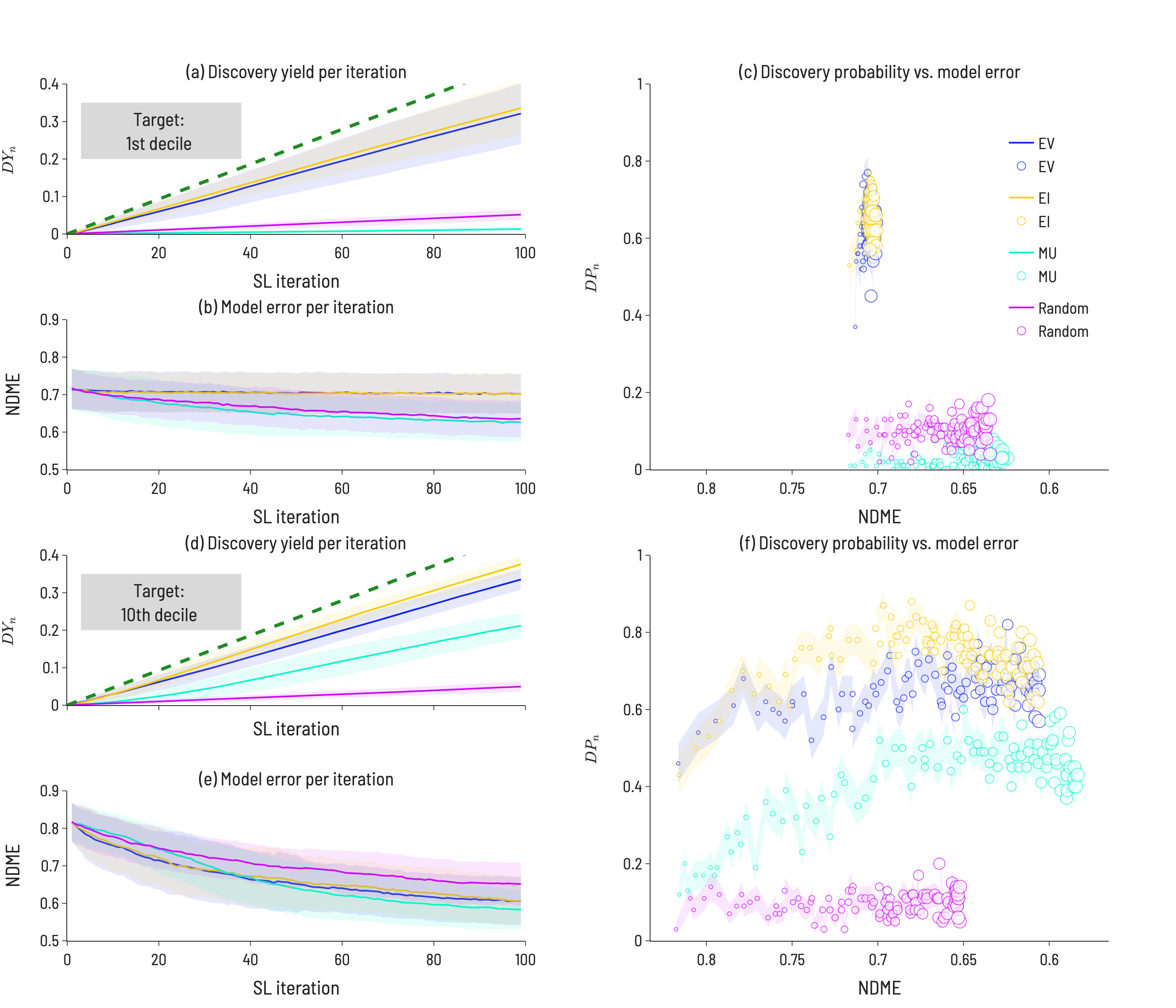}
    \caption{SL performance metrics for identifying (a-c) near-zero band gap materials (target = 1st decile) and (d-f) large band gap materials (target = 10th decile). (a) Discovery Yield ($DY(i)$) as a function of SL iteration for first decile materials (b) NDME as a function of SL iteration for first decile materials (c) Discovery probability ($DP(i)$) vs. NDME, where marker size grows with SL iteration number. $DY(i)$ is measure of how many target compounds have been identified after $i$ SL iterations and $DP(i)$ is a measure of the likelihood of identifying a target compound at SL iteration $i$. The error bars (denoted by the transparent colored bands) shown in the discovery yield and model error plots (panels a,b,d,e) represent the standard deviation between trials and the error bars shown in the discovery probability plots (panels c,f) represent the standard error between trials ($n_{trials}$=100). The green dashed line in panels (a) and (d) indicate a perfect target selection strategy.} 
    \label{fig:bandgap_dp}
\end{figure}

While Figure~\ref{fig:bandgap_heatmap} illustrates the performance of sequential learning in discovering a target number of compounds, we now consider an alternative scenario: a fixed number of sequential learning iterations.
In Figure~\ref{fig:bandgap_dp}, we set $n_{iter}$=100, and calculate performance metrics across the 100 iterations for the 1st and 10th decile target ranges.
This approach simulates the situation where a certain number of experiments can be performed and the optimum design strategy needs to be identified.
Interestingly, we find that no single acquisition function performs best in all cases and, for some targets (e.g., 1st decile materials), strategies that identify many target materials do not significantly reduce model error (relative to less performant strategies).
In Figure~\ref{fig:bandgap_dp}, the Discovery Yield ($DY_{i}$, $i$=1-100) is shown for 1st decile (panel a) and 10th decile (panel d) compounds, providing insight into the relative success in discovering target compounds using different acquisition functions.
Panels (b) and (e) show the NDME as a function of SL iteration. 
For each acquisition function, the number of target compounds found increases with SL iteration and the NDME decreases with SL iteration.
However, the rate at which these performance improvements occur depends on the acquisition function, the target range, and SL iteration number.
Therefore, the rate of successful target compound identification (represented by the Discovery Probability ($DP_{i}$, $i$=1-100) in panels (c) and (f)) is plotted against NDME to directly compare the relative improvements to materials discovery and model accuracy.
For both target ranges, EI and EV find the most target compounds (i.e., have the largest $DY_{100}$ and $DP_{100}$ values) of the four acquisition functions.
However, for model improvement, the best acquisition function depends on the target range, with MU as the most efficient at reducing NDME for the 1st decile, but EI and EV demonstrating superior performance for the 10th decile.
For the 1st decile, these performance metrics indicate that a decision must be made between ability to discover target compounds and model accuracy, as no single acquisition function does best in improving both.
In contrast, for the 10th decile, EI happens to outperform the other functions along both dimensions.
These results highlight the importance of assessing SL performance with metrics related to discovery of target compounds, rather than just model accuracy, as the two sets of metrics are not necessarily correlated with one another.

\subsection{Thermoelectrics}
We split thermoelectric materials design optimization into two sets of performance metrics: ZT and the pair of $\sigma_{E0}$ and thermal conductivity ($\kappa_{total}$).
While ZT is a traditional figure of merit for thermoelectric materials, it is often not indicative of the actual performance of a thermoelectric device due to large temperature variation of ZT and poor thermoelectric self compatibility across the temperature range of interest~\cite{snyder2006thermoelectric, snyder2017figure}.
To account for this, we also model $\sigma_{E0}$, a measure of the electronic mobility of a material and, hence, a material's potential to be a good thermoelectric.
$\sigma_{E0}$ is defined by Kang and Snyder~\cite{kang2017charge} as the prefactor, or transport coefficient, for a generalized transport model for conducting polymers or semiconductors based on Boltzmann transport equations (details of this calculation are provided in the supplementary information).
$\sigma_{E0}$ is an intrinsic property of a host matrix, independent of doping and thermal conductivity, and, therefore, can identify sub-optimally doped materials as promising thermoelectrics.
Since $\sigma_{E0}$ does not provide information about thermal conductivity, thermal conductivity is included in the analysis as well.
We also focus on 111-type compounds, as defined in Section \ref{datasets}.
As the Starrydata2 database does not contain structural information for all records, we cannot label “111-type” compounds as half-Heuslers, although we suspect that many such compounds in our dataset would indeed crystallize in the half-Heusler structure given the research interest surrounding this class of materials in the thermoelectric community~\cite{huang2016recent}.
Half-Heusler materials typically have high power factors but suffer from high thermal conductivity~\cite{uher1999dt}, providing opportunities for optimization through solid-solution alloying~\cite{xie2013beneficial}.

\begin{figure}
    \centering
    \includegraphics[width=\textwidth]{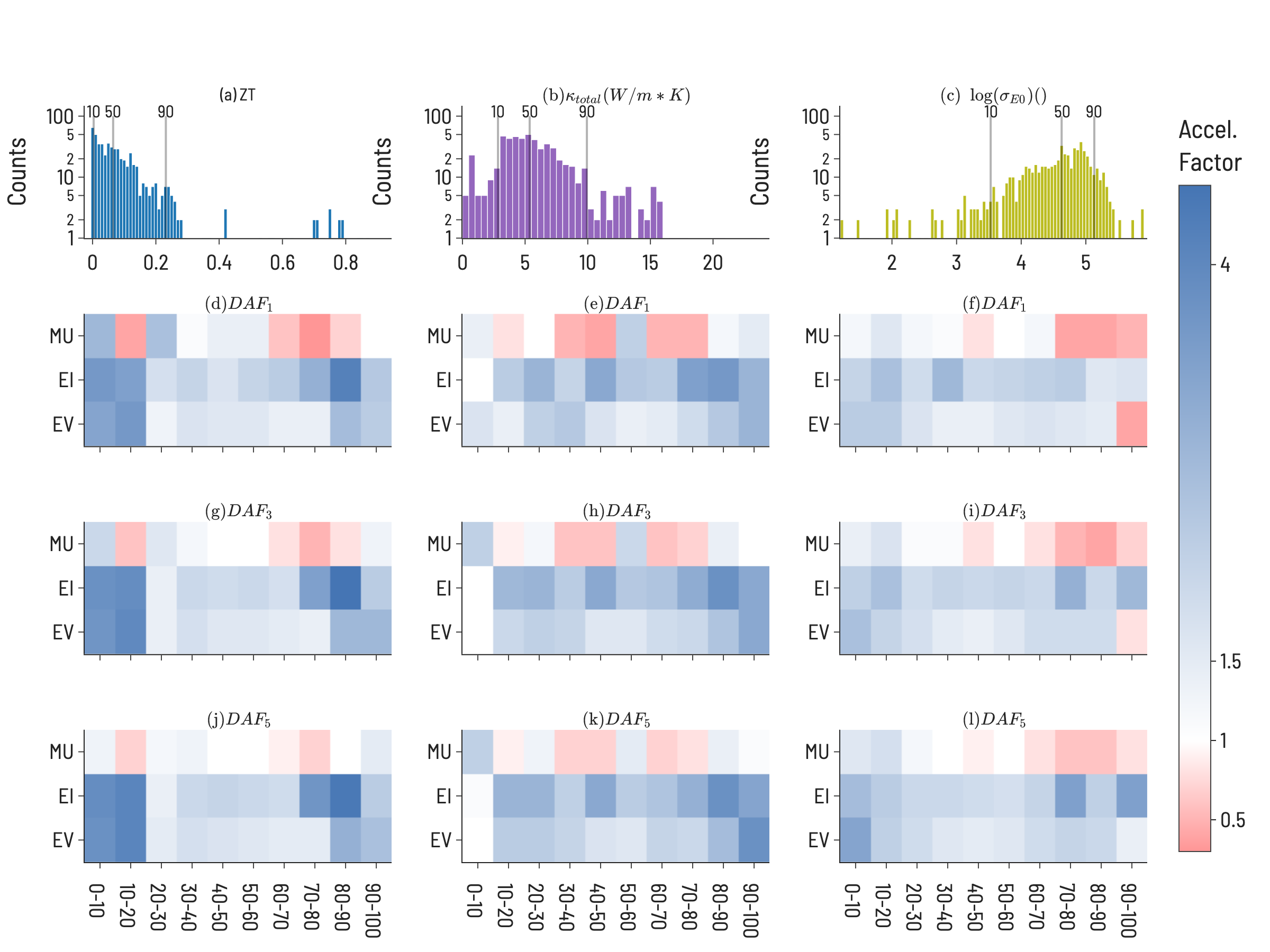}
    \caption{(a-c) Thermoelectric (TE) property dataset distributions and (d-l) Acceleration Factor ($DAF_n$) to first 1, 3, and 5 discoveries for TE properties of 111-type compounds. The top row shows the distribution of property values (ZT, $\kappa_{total}$, log($\sigma_{E0}$)) with the 10, 50, and 90 decile lines shown. A diverging color scheme is used to indicate performance (blue = the target compound(s) is(are) identified faster than RS, red = the acquisition function is outperformed by RS for the given target). DAF values were averaged over all trials ($n_{trials}$=100).} 
    \label{fig:te_heatmap}
\end{figure}

Histograms of the 111-type thermoelectric properties (ZT, $\kappa_{total}$, and log($\sigma_{E0}$)) in our dataset are shown in Figure~\ref{fig:te_heatmap}(a-c), illustrating a unique distribution for each property.
In general, small values of ZT are most plentiful; $\kappa_{total}$ exhibits a broad peak at small and intermediate values; and log($\sigma_{E0}$) has a peak at large values with negative skew.
The Discovery acceleration factor to identify  1, 3, or 5 target materials for the three thermoelectric properties are shown in Figure~\ref{fig:te_heatmap}.
Notably, we observe that it is more difficult to identify high-performing materials (large ZT, large log($\sigma_{E0}$), and small $\kappa_{total}$) than low-performing materials across all three properties.
EI and EV are almost always faster at identifying target materials than RS, with notable exceptions being large log($\sigma_{E0}$) and small $\kappa_{total}$.
This may indicate the difficulty of finding new high-performance thermoelectric materials using these design metrics.
By increasing the number of targets, EI and EV tend to improve relative to RS, although a few exceptions to this trend are apparent.
For example, when targeting $\kappa_{total}$ 1st decile materials, only MU exhibits improvement over RS when identifying 3 or 5 target materials and offers greater performance than EV or EI.
This is particularly interesting as MU does not perform well for high-performing ZT or log($\sigma_{E0}$) targets.
While EI appears to be a strong default acquisition function across properties, our results here suggest that an opportunity exists to tailor acquisition functions to specific problems, or to tune acquisition functions on the fly during SL processes.

\begin{figure}
    \centering
    \includegraphics[width=\textwidth]{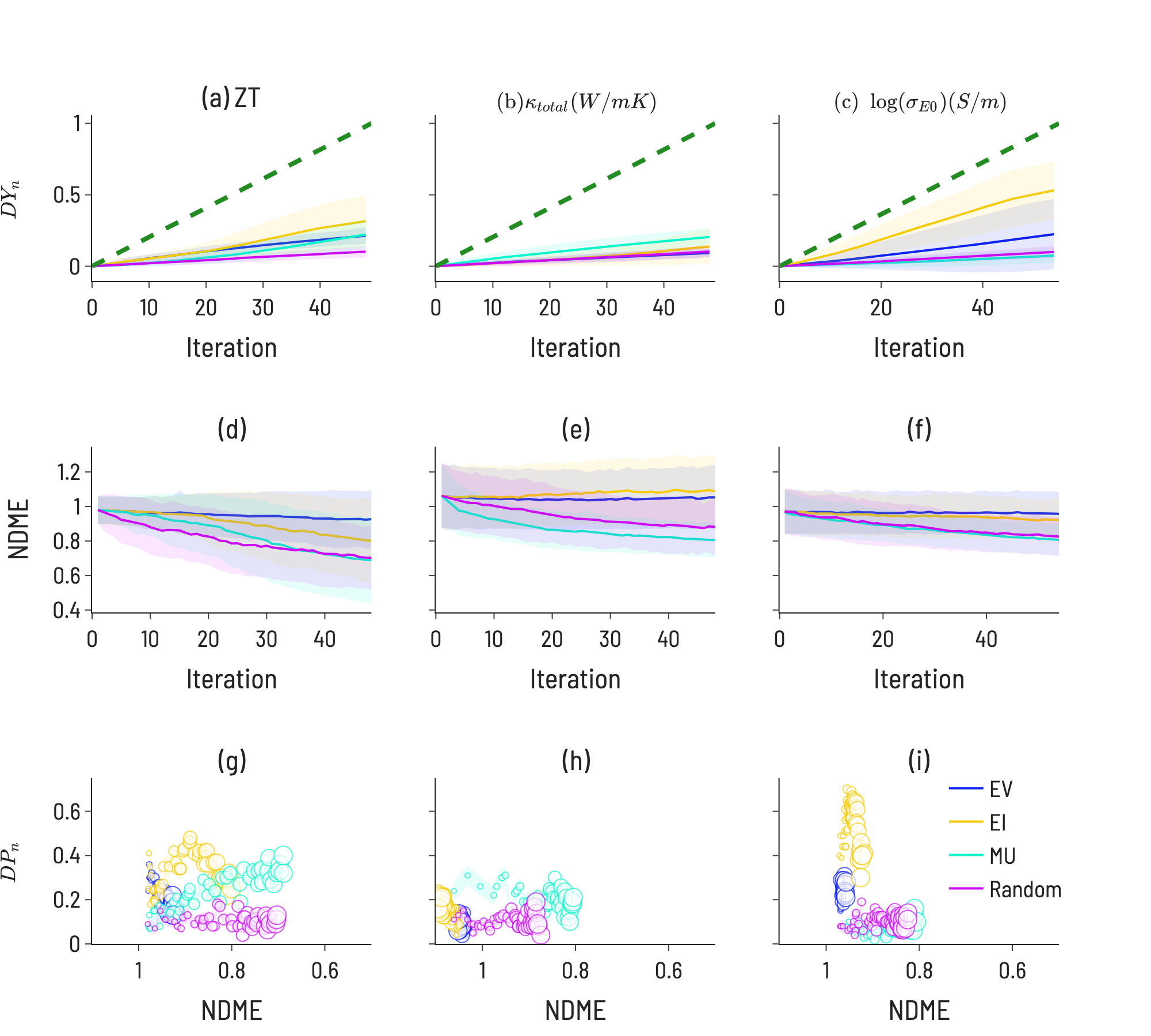}
    \caption{SL performance metrics for thermoelectric properties of 111-type compounds, (10th decile ZT (a,d,g), 1st decile $\kappa_{total}$ (b,e,h), and 10th decile $\sigma_{E0}$ (c,f,i). Discovery yield ($DY(i)$) is shown for all three properties over 100 SL iterations. Discovery probability ($DP(i)$) is shown for SL iteration equal to the number of target materials in the range (10\% of the dataset) (marker size increases with SL iteration number). The error bars (denoted by the transparent colored bands) represent the standard deviation between trials ($n_{trials}$=100).} 
    \label{fig:te_dp}
\end{figure}

The results of simulating SL to find high performing thermoelectric materials (10th decile ZT, 1st decile $\kappa_{total}$, and 10th decile log($\sigma_{E0}$)) for 100 iterations are shown in Figure~\ref{fig:te_dp}.
For ZT, EI and MU appear to be the highest-performing design strategies, with better ability to improve both model accuracy and target identification over EV, indicating a need for model uncertainty information in ZT design.
In contrast, a choice must be made for log($\sigma_{E0}$) design to either identify high-performance materials (with EI) or improve model accuracy (with MU).
$\kappa_{total}$ has the unique behavior that MU performs best for both improving model accuracy and finding target compounds, indicating a particular difficulty in modeling thermal conductivity.
Amongst the three properties, $\kappa_{total}$ has the lowest Discovery Yield and highest NDMEs.
A further observation is that, across many use cases we examine here, EV is not the most performant acquisition function.
However, an important use case for EV is apparent in Figure~\ref{fig:te_heatmap}: discovery at early rounds SL iterations.
If resource constraints limit a materials discovery effort to only a few candidate evaluations, EV may be the appropriate choice of acquisition function for certain tasks (e.g., finding low thermal conductivity materials).

\subsubsection{Role of experimental metadata in sequential learning}
\begin{figure}
    \centering
    \includegraphics[width=0.8\textwidth]{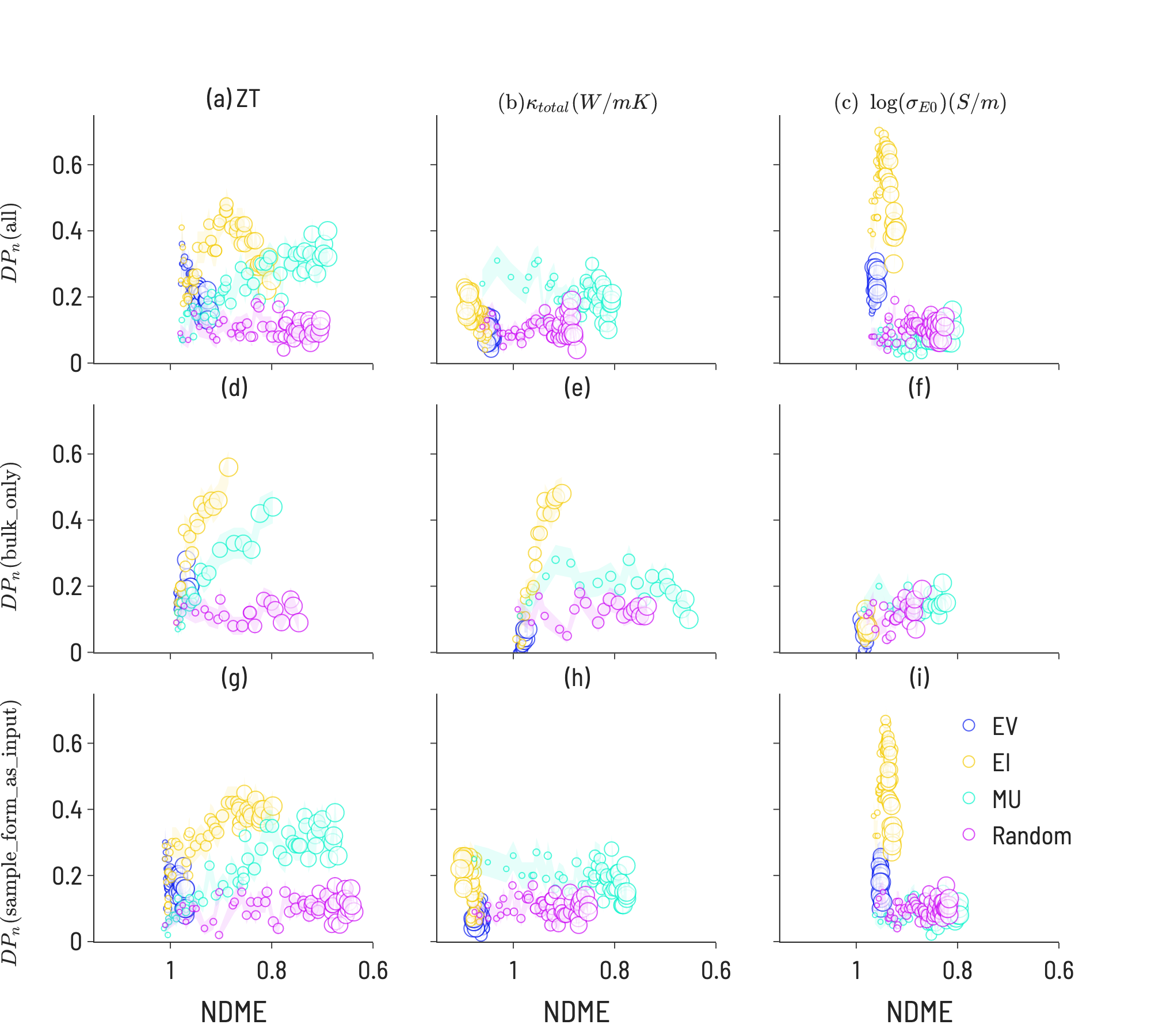}
    \caption{Discovery probability plots for TE properties in datasets with varying metadata (marker size increases with SL iteration number): (a-c) all 111-type samples (presented Figure~\ref{fig:te_dp}, (d-f) bulk only samples, (g-i) adding sample form as an input to ML model training.) The error bars (denoted by the transparent colored bands) represent the standard deviation between trials ($n_{trials}$=100).} 
    \label{fig:te_dp_compare}
\end{figure}

To assess the effect of metadata on SL performance, we compare thermoelectric SL metrics for different sets of inputs in Figure~\ref{fig:te_dp_compare}. 
As has been shown previously~\cite{murdock2020domain}, supplying additional domain knowledge can further boost the performance of ML models.
To that end, we suspect that the manner of sample preparation (e.g., thin film vs. bulk) may impact the resulting TE properties (i.e., the same composition in two different sample forms will have different properties).
As all results up to now have not included this information, we hypothesize that adding sample form as input to ML models or limiting to just bulk data will affect SL performance. 
To test this, we perform two alternative SL workflows utilizing labels found in StarryData: (1) filtering TE datasets to include only samples marked as "bulk" and (2) incorporating an additional feature, "sample form", as an input.
As the majority of samples were not labeled with sample form, when filtering on samples labeled as "bulk, the dataset sizes for ZT, $\kappa_{total}$, and $\sigma_{E0}$ were reduced from 626, 618, and 705 to 285, 274, and 301, respectively.
The addition of "sample form" as a feature resulted in dataset sizes for ZT, $\kappa_{total}$, and $\sigma_{E0}$ of 671, 654, and 749, respectively.
The slight difference in dataset size (relative to the original workflow) is due to the samples with the same composition having different forms, which increases the number of rows in the dataset.
For example, there are 31 samples with the composition ZrNiSn.
Of these samples, 14 are denoted as "bulk" and the other 17 are unknown or not labeled.
This results in two records in the dataset: the mean property value of the 14 bulk samples and the mean property value of the 17 unknown samples.
In the original workflow, this would collapse to a single record: the mean property value of all 31 samples.

The results of the running the SL workflow with different amounts of metadata are presented in Figure~\ref{fig:te_dp_compare} along with the original workflow for comparison.
The most notable differences are seen between the bulk samples and all samples.
For $\kappa_{total}$, EI results in the largest $DP(i)$ when conducting the SL workflow on only bulk samples, while MU results in the largest $DP(i)$ when conducting the SL workflow on all samples or when including sample form as an input.
This finding correlates with experimental observations~\cite{bhattacharya2002grain}, which identified a microstructural effect on thermal conductivity for 111-type compounds and suggests that high-performance SL for $\kappa_{total}$ is only possible with a high-quality description of microstructure.
Further, this illustrates that microstructural differences have a direct effect on SL outcomes and performance, highlighting the importance of integrating domain knowledge into ML material representations.
Additionally, we note that EV performs relatively poorly in both workflows, indicating the benefits of including uncertainty estimates with ML predictions.
For log($\sigma_{E0}$), EI is less performant when conducting the SL workflow on only bulk samples, suggesting that changing the shape of the target property distribution can have consequences for performance.
It may also be important to examine the behavior of SL for different quantities of training data.
Including "sample form" as an input did not change the SL results meaningfully when compared to a composition-only model.
This is likely due to a large number of samples having no "sample form" label, as over 50\% of samples had empty or unknown entries for this label.

\section{Discussion}
When determining which acquisition function to use in sequential learning, there is a fundamental trade-off between identifying high-performing materials and improving conventional model accuracy metrics.
As shown in Figures \ref{fig:bandgap_dp} and \ref{fig:te_dp}, model accuracy alone is not a reliable indicator of a model’s ability to discover target materials.
While both NDME and $DP(i)$ can improve simultaneously, in several of the cases herein (notably 1st decile band gap (Figure~\ref{fig:bandgap_dp}f) and 10th decile log($\sigma_{E0}$) (Figure~\ref{fig:te_dp}i)), a decrease in NDME is not associated with an improvement in finding target materials (and vice versa).
Additionally, these particular results suggest that models with NDME$\approx$1 are able to discover materials in the target regions, illustrating that such models can still be viable for materials discovery applications.
A decoupling of model error and the ability of that model to identify high-performance materials is consistent with the results of Rohr et al.~\cite{rohr2020benchmarking}, who showed that it is possible to identify high-performing catalysts even when model performance (as measured by MAE) is not significantly improved through SL.

Our results demonstrate that SL performance depends sensitively on the target range within the property distribution, how many SL iterations are allowed, the number of targets to discover, and the incorporation of uncertainty estimates in the SL acquisition functions.
The nature of the design problem has a large effect on the possibility of SL success.
The heatmaps in Figures \ref{fig:bandgap_heatmap} and \ref{fig:te_heatmap} indicate a large variation in $DAF_n$ to find targets for a given property amongst the different target ranges.
For instance, finding one material within the 10th decile range of ZT values is much more difficult than finding one in the 9th decile range.
This is likely dependent on the distribution of materials property values in the dataset, particularly in the feature space where the models are trained.
Such heatmaps can be used to optimize model architecture and design problems for the particular distribution of a given dataset.
For instance, it appears particularly difficult to identify 1st decile $\kappa_{total}$ materials (Figure~\ref{fig:te_heatmap} middle column), at least without knowledge of microstructure (as suggested in Figure~\ref{fig:te_dp_compare}).
Thus, one possible screening strategy could be to filter out materials predicted to lie in the 2nd-10th deciles and focus on the remaining candidates.

Simulated sequential learning on existing data can inform the selection of an acquisition function for ongoing SL.
Based on $DAF_n$ heatmaps (such as those in Figures \ref{fig:bandgap_heatmap} and \ref{fig:te_heatmap}), the difficulty of a particular target range can be assessed. 
$DY(i)$ and $DP(i)$ plots, such as those in Figures \ref{fig:bandgap_dp} and \ref{fig:te_dp}), can be used to estimate the trade-off between materials discovery and model accuracy at early and later iterations of SL.
The optimal discovery approach depends on the particular materials design goal at hand and no single acquisition function necessarily performs best in all cases, although EI appears to be a robust default choice.






\section{Summary and Conclusions}
We present a framework and set of metrics for evaluating the performance of sequential learning across materials design problems, and we find that SL performance varies greatly depending on the particulars of a design problem.
For example, configurations that quickly find the first target material may have limited success in finding the second and third.
Similarly, those configurations which are slow to improve model accuracy may be quick to find the most materials in a given target range.
Moreover, for situations where SL strategies for identifying high performing materials are lacking, strategies for identifying materials in other performance regimes may still be successful, suggesting alternative design strategies (e.g., utilizing ML models to indicate materials or regions of design space to avoid).
Such findings illustrate the importance of testing many sequential learning configurations on new datasets.
The two case studies reported here highlight the importance of selecting an appropriate optimization strategy based on goals of the specific SL effort, and dynamical metrics such as Discovery Yield and Discovery Probably provide better measure of model performance for discovering new materials than static metrics like RMSE.

\section{Data availability}
The datasets curated for this article are made available at \href{https://github.com/CitrineInformatics-ERD-public/sl_discovery}{https://github.com/CitrineInformatics-ERD-public/sl\_discovery}.

\section{Code availability}
The scripts for the sequential learning workflow for this article are made available at \href{https://github.com/CitrineInformatics-ERD-public/sl_discovery}{https://github.com/CitrineInformatics-ERD-public/sl\_discovery}.

\section{Conflicts of Interest}
There are no conflicts to declare.

\section{Acknowledgements}
This work was supported by the Department of Energy, Energy Efficient and Renewable Energy program under contract DE-AC02-76SF00515. We gratefully acknowledge scientific discussions with Vinay Hegde, Andrew Lee, Ramya Gurunathan, and Suchismita Sarker. 
\bibliographystyle{unsrt}  
\bibliography{references.bib}

\newpage
\section{Supplementary Information}

\subsection{Starrydata ingestion pipeline}

\begin{figure}
    \centering
    \includegraphics[width=\textwidth]{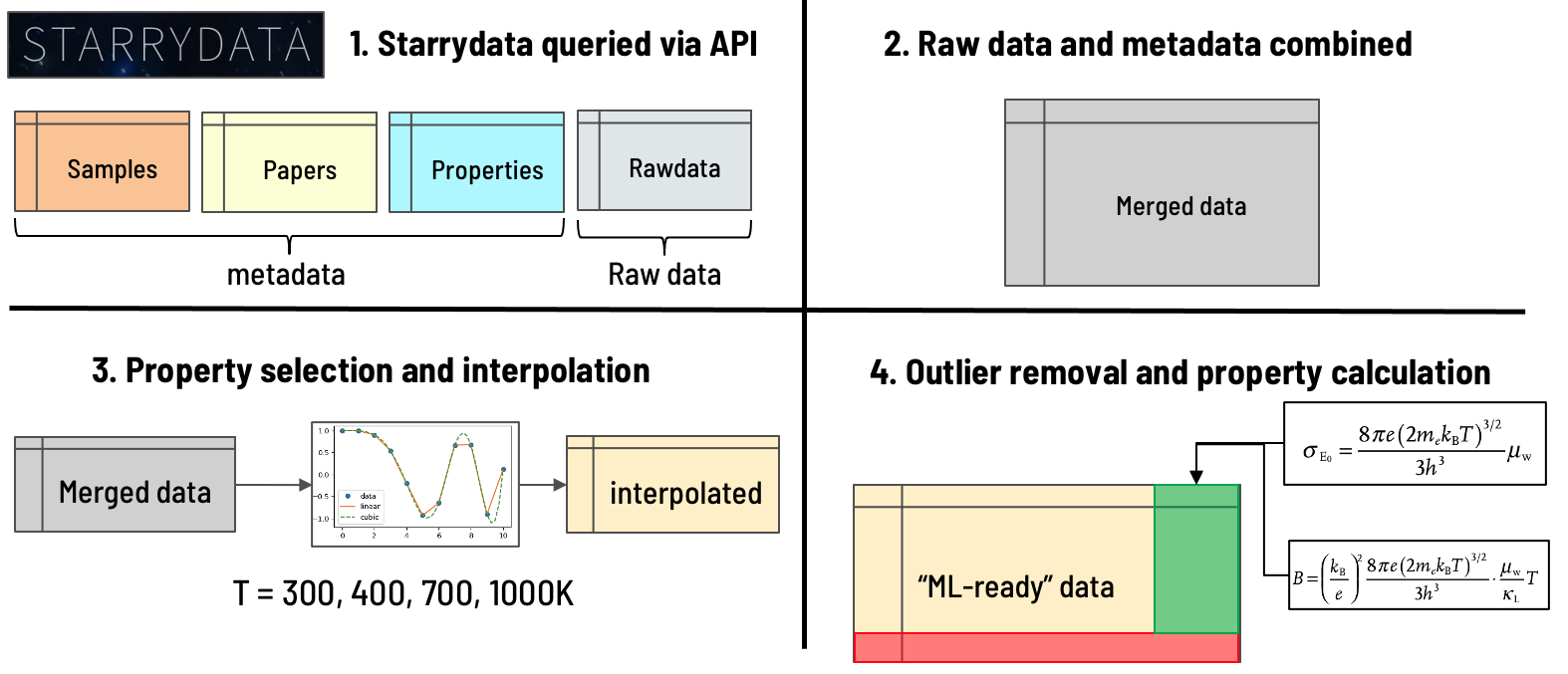}
    \caption{Starrydata ingestion pipeline. Raw data is queried via the starrydata api and merged into a single table. Then, Data points are interpolated at standard temperatures, thermoelectric figures of merit (e.g., $\sigma_{E0}$) are calculated, and physically-intuitive filters are applied to remove erroneous data points.} 
    \label{fig:sd-ingest}
\end{figure}

As illustrated in Figure~\ref{fig:sd-ingest}, the Starrydata ingestion pipeline consists of four processes.
First, data from the Starrydata2 database is queried via the API.
Specifically, a generalized query for any sample containing an element from the periodic table was constructed, and queried samples were cached in a local directory (note, samples not containing any elements were not returned).
Queried samples were then converted to individual Pandas dataframes (raw, sample, paper, property) as shown in the Starrydata documentation~\cite{starryweb2021}.
Second, the raw data was combined with data from sample, paper, and property dataframes.
This step generates one easily viewable table with all important linkages (process-composition-property).
Third, since many properties of interest are recorded as a function of temperature, these properties are interpolated at defined temperatures (T = 300K).
This allows for direct comparison of samples and their properties at defined temperatures.
Fourth, thermoelectric figures of merit (e.g., $\sigma_{E0}$~\cite{kang2017charge}) are calculated and subsequently filtered to identify erroneous data points.
To denote a compound as being "111-type", we first ensure a composition has 3 or more elements and that all elements are either a metal or metalloid.
Then, if the sum of the stoichiometry is between 2.9 and 3.1 and no single element has greater than 0.4 atomic fraction, the compound was denoted as being "111-type". 

After querying for Starrydata records, it was noted that many records are sparse, only having a few properties reported for a given composition.
This is likely due to the nature of the way experimental properties for TEs are reported and how they are extracted by the Starrydata team.
When applicable, TE properties are calculated from extracted (interpolated) values using empirical relationships.
For example, if Power Factor (PF) is not reported for a sample but Seebeck and electrical conductivity are reported for that sample, PF is calculated via: \(PF = S^2*\sigma\).
These calculated values are then combined with values extracted (interpolated) from the Starrydata2 database.
In instances where a sample has both an extracted and calculated value for a property, the extracted value was used.
The number of properties extracted and calculated for samples at room-temperature are shown in Table \ref{table:prop_table}.
All property calculations are listed below:

\begin{enumerate}
    \item \(\sigma = \rho^{-1}\) ($\sigma$ = electrical conductivity, $\rho$ = electrical resistivity)
    \item \(PF = S^2*\sigma\)  (PF = Power Factor, S = Seebeck coefficient)
    \item \(ZT = PF * T * \kappa_{total}^{-1}\)  (T = Temperature, $\kappa_{total}$ = Total thermal conductivity)
    \item $\sigma_{E0} \rightarrow e(2m_{e}\kappa_{B}T)^{3/2}/2\pi^2\hbar^3 * \mu_{0}(m^{*}/m_{e})^{3/2}$. ($\sigma_{E0}$ = transport model prefactor,  $\mu_{0}(m^{*}/m_{e})^{3/2}$ = weighted mobility)
\end{enumerate}


\begin{table}
\centering
\caption{Physically-relevant ranges for properties of interest. Property values outside of these ranges were filtered out by the SL pipeline.}
\label{table:prop_ranges}
\begin{tabular}{llll}
\toprule
                               Property &      units & minimum value & maximum value \\
\midrule
                Seebeck coefficient (S) &      $V/K$ &        -0.005 &         0.005 \\
     Electrical conductivity ($\sigma$) &      $S/m$ &             0 &      10000000 \\
Thermal conductivity ($\kappa_{total}$) &   $W/(mK)$ &             0 &           100 \\
             Power factor (S$^2\sigma$) & $W/(mK^2)$ &             0 &            10 \\
                   Figure of merit (ZT) &          - &             0 &             3 \\
                        Temperature (K) &          K &           200 &          1200 \\
  Transport coefficient ($\sigma_{E0}$) &      $S/m$ &             0 &      10000000 \\
\bottomrule
\end{tabular}
\end{table}

\begin{table}
\centering
\caption{Thermoelectric properties extracted from the Starrydata2 database. Raw data, interpolated, calculated, and final refer to distinct caches at points in the data ingestion pipeline 111-type refers to records that are labeled as 111-type by our composition classifier.}
\label{table:prop_table}
\begin{tabular}{lllrrrrr}
\toprule
Property ID &                              Property &      Units &  raw data &  interpolated &  calculated &  final &  111-type \\
\midrule
          2 &                   Seebeck coefficient &        V/K &    498527 &         21508 &           0 &  17315 &       986 \\
          3 &               Electrical conductivity &      $S/m$ &    184924 &          9931 &       10366 &  16438 &       970 \\
          4 &                  Thermal conductivity &   $W/(mK)$ &    276597 &         15508 &           0 &  12785 &       789 \\
          5 &                Electrical resistivity & $\Omega m$ &    324818 &         10399 &           0 &   8495 &       462 \\
          6 &                          Power factor & $W/(mK^2)$ &    184900 &         10570 &        8529 &  15437 &       913 \\
          8 &                                    ZT &          - &    221091 &         13730 &        1713 &  12794 &       808 \\
sigma\_E\_0 & Transport coefficient ($\sigma_{E0}$) &      $S/m$ &         0 &             0 &       18181 &  14742 &       889 \\
\bottomrule
\end{tabular}
\end{table}

Table~\ref{table:prop_table} highlights the number of records at distinct point in the data processing pipeline.
Raw data refers to the number of data points for a particular property obtained from the initial query of the database.
The initial query returns records with dense property curves (e.g. Seebeck as a function of temperature).
To standardize the temperature at which properties are compared, property-temperature curves were interpolated at 300K; after filtering for curves with a physically-relevant temperature range (shown in table~\ref{table:prop_ranges}). 
This reduced the amount of data per record to the number of data points shown in the "interpolated" column.
As previously described, the "calculated" column shows the number of data points for a particular property that were calculated from interpolated values using empirical relationships.
The "final" dataset shows a filtered view of both calculated and interpolated values based on the physically-relevant properties filters shown in Table~\ref{table:prop_ranges}.
The "111-type" column shows the same data points from "final" for only compounds that agree with our 111-type composition classifier (described in previous paragraph).
The "111-type" dataset was used as an input to the SL pipeline resulting in the record counts shown in Section~\ref{datasets} (after indexing on chemical formula and filtering $ZT_{max}$ = 2).

\subsection{Acquisition function notation}

When making predictions using random forests with uncertainty estimates, a prediction can be represented as a predicted distribution (rather than a single value) where the width of the distribution is correlated with the uncertainty of a prediction. 

We note that previous work on sequential learning\cite{ling2017high} used alternative definitions for acquisition functions:
\begin{enumerate}
    \item Maximum Expected Improvement (MEI) = The candidate with the highest (or lowest, for minimization) target value.
    \item Maximum Likelihood of Improvement (MLI) = The candidate with the highest (or lowest, for minimization) target value when including prediction uncertainty.
    \item Maximum Uncertainty (MU) = The candidate with the greatest prediction uncertainty, entirely independent of its expected value.
    \item Random search (RS) = A candidate is selected at random from the pool.
\end{enumerate}

Herein, we use definitions that are more consistent with other literature:
\begin{enumerate}
    \item Expected Value (EV) = The candidate with a predicted distribution mean value closest to the target value. EV strictly favors values closer to the target without considering uncertainty.
    \item Expected Improvement (EI) = The candidate with a predicted distribution curve that has the highest probability of intersecting with the target value.
    \item Maximum Uncertainty (MU) = The candidate with the greatest prediction uncertainty, entirely independent of its expected value.
    \item Random search (RS) = A candidate is selected at random from the pool.
\end{enumerate}

Note: Expected Value (EV) is consistent with the previously defined Maximum Expected Improvement (MEI) and Expected Improvement (EI) is consistent with the previously defined Maximum Likelihood of Improvement (MLI).

\end{document}